\documentstyle[12pt]{article}
\begin{document}
\begin{center}
{\bf   PARTICLE CREATION IN ANISOTROPICALLY EXPANDING UNIVERSE }

{ P.K.SURESH  }{\footnote{email:pkssp@uohyd.ernet.in}} \\[.2cm]

{\it School of Physics, University of Hyderabad. Hyderabad 500 
046. India.}  
\end{center}
\begin{abstract}
~~Using squeezed vacuum states formalism of quantum optics,
an  approximate  solution to the semiclassical Einstein equation is obtained in Bianchi type-I 
universe. The phenomena of nonclassical
 particle creation is also examined in the anisotropic background cosmology.
\end{abstract}

{\small {\emph{Keywords}: Bianchi type-I, cosmology, Einstein equation, particle creation, squeezed vacuum. }\\
{\small PACS numbers:  42.50.-p, 42.50.Dv, 98.80.k, 98.80.Cq, 98.80.Qc}
 }

{\bf 1. Introduction}

The present universe in its over all structures
 seems to be spatially homogeneous and isotropic but there 
are reasons to believe that it has not been so in all its 
evolution and that inhomogeneities and  anisotropies might have 
played an important role in the early universe $^{1,2}$ . The 
isotropic model is adequate enough for the description of the 
later stages of evolution of the universe  but this does not mean 
that the model  equally suite for the description of very early 
stages of the evolution of the universe, especially near the  singularity $^{2}$. 
Also the most general solution of the problem  of gravitational 
collapse turn to be locally anisotropic near the singularity 
$^{3-5}$. Cosmological solutions of Einstein's general relativity 
are also known in which the expansion can be anisotropic at first, 
near the singularity, and later the expansion became 
isotropic. Also, to
 avoid postulating specific initial 
conditions as well  as  the existence of particle horizon in 
isotropic models, attempts have been made through the  study  of 
inhomogeneous and anisotropic models of the universe. Among the 
anisotropic cosmological models, Bianchi type -I universe is the 
simplest one. In this model the metric considered as  spatially 
homogeneous   and  possibly anisotropic. In contrast to the 
Friedmann-Robertson-Walker (FRW) metric,
 Bianchi type-I metric has three scale factors which
 evolve differently in their respective direction.
 Therefor the expansion in this model could be 
considered as anisotropic expansion. Interests in such models 
have  been  received much attention $^{1,3-7}$. Huang has 
considered the fate of symmetry in  Bianchi type - I cosmology 
using adiabatic approximation for massless field with arbitrary 
coupling to gravity  $^{8}$. Futamase has studied the effective 
potential in Bianchi type - I cosmology  $^{9}$. Berkin has 
examined the effective potential in Bianchi type -I universe, for 
scalar field having arbitrary mass and coupling to gravity  
$^{10}$. These studies show that 
Bianchi type -I cosmological model may be useful to understand the very early 
universe. Anisotropic models of the universe which become 
isotropic  have been  considered  several times
$^{11}$. These motivate the study of anisotropic background 
cosmological models with scalar field possess the advantage of 
FRW model, and to analysis the possibility of 
its approach to the isotropic model with  accuracy required 
by observations $^{12}$. With this aim Bianchi type -I universe 
has studied and the relevant equation for which extreme close in 
the form of isotropic case examined. 
Form anisotropic to isotropic transition a 
damping mechanism is required.  One of the efficient damping 
mechanisms could be due to the particle creation in  anisotropic 
models. Therefore it would be useful to examine the particle creation in the anisotropic cosmological model with nonclassical inflaton, which could expect to produce sufficient particles to bring isotropy during the evolution process of the universe.

In this paper we study a homogeneous massive scalar field
 (inflaton), minimally coupled to gravity, in a  spatially homogeneous and anisotropic background 
metric. The inflaton under our 
consideration can be quantized and represent in squeezed vacuum states, 
 hence an approximate solution to the semiclassical Einstein equation and 
the phenomenon of nonclassical particle creation  can be  
examined in semiclassical theory of gravity.

Throughout the  paper, we follow the units $ c=G=\hbar=1$.

{\bf 2. Inflaton in Bianchi type -I metric}

The Bianchi type- I model 
is an anisotropic generalization of the FRW model with Euclidean 
spatial geometry. The Bianchi type -I metric can be considered as spatially homogeneous and anisotropic and is given by:
\begin{eqnarray}
ds^2=   dt^2 - {\cal{S}}_{1}^2(t) d{x _{1}}^2- {\cal{S}}_{2}^2(t) 
d{x _{2}}^2 -{\cal{S}}_{3}^2(t) d{x _{3}}^2,
\end{eqnarray}
where  ${\cal{S}}_1(t)$, ${\cal{S}}_2(t)$ and  ${\cal{S}}_3(t)$ 
are the scale factors in three spatial directions. 
 Which are representing the size of the universe in 
their respective  direction.  The three scale factors  ${\cal{S}}_1(t)$, 
${\cal{S}}_2(t)$ and  ${\cal{S}}_3(t)$ are determined via 
Einstein's equations.  

In the background metric (1), consider  an inflaton, minimally 
coupled to gravity, satisfy the equation:
\begin{eqnarray}
\left (  g^{\mu \nu} \nabla _{\mu} \nabla _{\nu}- m^2 \right ) 
\phi(x,t)=0,
\end{eqnarray}
where $\nabla_{\mu} $ is the covariant derivative. The Lagrangian  
density for the inflaton field, $\phi$, is given by:
\begin{equation}
{\cal{L}}= - {1 \over 2} {\sqrt{-g}}\left ( g^{\alpha \beta} 
\partial_{\alpha} \phi \partial_{\beta} \phi + m^2 \phi^2 \right ). 
\end{equation}
 Since 
the inflaton is homogeneous,ie.; $\phi(x,t)=\phi(t) $, its classical equation of motion  for the metric (1) can be written as 
\begin{eqnarray}
{\ddot{\phi}}(t) + \sum_{i=1} ^3 \left ( { 
{\dot{{\cal{S}}_i}}(t)\over {\cal{S}}_i(t)} \right ) 
{\dot{\phi}}(t)+ m^2 \phi(t)=0. 
\end{eqnarray}
In the present context (4) is the classical equation of motion 
for the inflaton for the metric (1).
 
For the metric (1), the purely temporal component of the 
classical gravity is now the classical Einstein equation and 
can be written as:
\begin{eqnarray}
{\dot{{\cal{S}}_1}(t)\over {\cal{S}}_1(t)} 
{\dot{{\cal{S}}_2}(t)\over {\cal{S}}_2(t)} + 
{\dot{{\cal{S}}_2}(t)\over {\cal{S}}_2(t)} 
{\dot{{\cal{S}}_3}(t)\over {\cal{S}}_3(t)} + 
{\dot{{\cal{S}}_1}(t)\over {\cal{S}}_1(t)} 
{\dot{{\cal{S}}_3}(t)\over {\cal{S}}_3(t)} =
 {8 \pi  T_{00}\over  {\cal{S}}_1(t) {\cal{S}}_2(t) 
{\cal{S}}_3(t)},
\end{eqnarray}
where
\begin{eqnarray}
T_{00}=  {\cal{S}}_1(t) {\cal{S}}_2(t) {\cal{S}}_3(t)
 \left ( {{\dot{\phi}}^2 \over 2} + m^2 { \phi^2(t) \over 2}\right 
 ),
\end{eqnarray}
is the energy density of the inflaton. In  cosmological context, 
the classical Einstein equation (5), means  that the Hubble constants 
 $\left ( H_{i}= { {\dot{{\cal{S}}_{i}}}(t)\over 
{\cal{S}}_{i}(t)} \right )$ are determined by the energy density 
of the dynamically evolving inflaton described by the classical 
equation of motion.

In order  to study the full quantum effects in a cosmological 
model, 
 both metric and matter  fields are to be treated quantum mechanically. 
 Since a consistent quantum theory of 
gravity is not available, in most of the cosmological models  the 
background metric  under consideration is taken as classical form 
and matter field treat as quantum mechanical. Such approximation 
of the Einstein equation is know as semiclassical approximation.

In semiclassical theory  Einstein equation takes the following 
form 
\begin{eqnarray}
G_{\mu \nu}= 8 \pi  \langle {\hat{T}}_{\mu \nu} \rangle,
\end{eqnarray}
where $G_{\mu \nu}= R_{\mu \nu}-{1\over2}g_{\mu \nu}R$ is the Einstein 
tensor and $\langle {\hat{T}}_{\mu \nu} \rangle$ is the 
expectation value of the energy-momentum tensor for a matter 
field in a suitable quantum state under consideration. In (7) the quantum field can be represented by a scalar field, 
$\phi$, and is governed by the time dependent Schrodinger equation 
\begin{eqnarray}
i{ \partial \over \partial 
t}\Psi(\phi,t)={\hat{H}}_{m}(\phi,t)\Psi(\phi,t).
\end{eqnarray}

Using the canonical quantization procedure, the scalar field can 
be quantized 
 by defining the momentum conjugate to $\phi$, as 
\begin{eqnarray}
 \pi_{\phi} = { \partial { \cal{L}} \over  \partial \dot{\phi} }.
\end{eqnarray}
Thus the  inflaton in the Bianchi type -I cosmological model, can 
be described by a time dependent harmonic oscillator with the Hamiltonian
\begin{eqnarray} 
H_m= {1 \over 2  {\cal{S}}_1(t) {\cal{S}}_2(t) 
{\cal{S}}_3(t)}\pi_{\phi}^2 + {  m^2  {\cal{S}}_1(t) 
{\cal{S}}_2(t) {\cal{S}}_3(t) \over 2} \phi^2(t).
\end{eqnarray}

The eigenstates of the Hamiltonian are the Fock states which  can 
be constructed by annihilation and creation operators in the 
following manner.
\begin{eqnarray}
\hat{a}(t)= \phi^*(t){\hat{\pi}}_{\phi}-
 {\cal{S}}_1(t) {\cal{S}}_2(t) {\cal{S}}_3(t) {\dot{\phi}}^*(t) \hat{\phi},  \\ \nonumber
\hat{a}^{\dagger}(t)=\phi(t){\hat{\pi}}_{\phi}- {\cal{S}}_1(t) 
{\cal{S}}_2(t) {\cal{S}}_3(t) \dot{\phi}(t) \hat{\phi}.
\end{eqnarray}
Thus  the Fock state of  Hamiltonian is  :
\begin{eqnarray}
\hat{a}^{\dagger}\hat{a} \mid n,\phi,t \rangle = n \mid n,\phi,t 
\rangle.
\end{eqnarray}

In the present context the semiclassical  Einstein equation takes 
following form:
\begin{eqnarray}
{\dot{{\cal{S}}_1}(t)\over {\cal{S}}_1(t)} 
{\dot{{\cal{S}}_2}(t)\over {\cal{S}}_2(t)} + 
{\dot{{\cal{S}}_2}(t)\over {\cal{S}}_2(t)} 
{\dot{{\cal{S}}_3}(t)\over {\cal{S}}_3(t)} + 
{\dot{{\cal{S}}_1}(t)\over {\cal{S}}_1(t)} 
{\dot{{\cal{S}}_3}(t)\over {\cal{S}}_3(t)} =   \frac{8 
\pi}{{\cal{S}}_1(t) {\cal{S}}_2(t) {\cal{S}}_3(t)} 
                \langle \hat{H} \rangle,
\end{eqnarray}
where $H$  is given by (10).

 {\bf 3. Particle Creation
in the anisotropic universe}

Most of the cosmological models  are based on a classical 
behaviour of the scalar field. Therefore, it is of interest to 
study the evolution of the
 system with the scalar field, which possess the nonclassical features. 
Recently squeezed states formalism of quantum
 optics$^{13}$ is found much useful to deal with many issues in cosmology$^{13-23}$.
Squeezed states are minimum uncertainty states and are obeying 
Heisenberg uncertainty principle. A squeezed state is generated 
by the action of squeezed operator on any coherent state and, in 
particular, on the vacuum state. Therefore a squeezed vacuum can 
be  defined $^{13}$ 
\begin{equation}
\mid {\cal{Z}} \rangle={\cal{Z}}(r,\varphi)\mid 0\rangle,
\end{equation}
where the squeezing operator,
\begin{equation}
{\cal{Z}}(r,\varphi)=\exp\left ( {r\over2}\left (e^{-i\varphi} 
a^2 - e^{i\varphi}{a^{\dagger}}^2 \right ) \right).
\end{equation}
In  (15) $r$ is the squeezing parameter which determines the 
strength of squeezing and $\varphi$ is the squeezing angle which 
determines the distribution between the conjugate variables and 
they take values 0$\leq r <   \infty$ and 
$-\pi\leq\varphi\leq\pi$. $ a$ and $ a^{\dag}$ are respectively known as annihilation and creation operators. When the squeezing operator acts on annihilation and creation operators lead the following results$^{13}$:
\begin{eqnarray}
{\cal{Z}}^{\dagger}a{\cal{Z}}=a\cosh r -a^{\dagger}e^{i\varphi}\sinh r ,\\
\nonumber 
{\cal{Z}}^{\dagger}a^{\dagger}{\cal{Z}}=a^{\dagger}\cosh r 
-ae^{-i\varphi}\sinh r.
\end{eqnarray}
In the case of squeezed states, the variance of the quadrate 
components are not equal but one component of the noise is always 
squeezed with respect to another.

Next, consider the Hamiltonian  of the semiclassical Einstein equation (13), whose 
 expectation value can be computed in squeezed vacuum 
state by replacing the number 
 state $ \mid n,\phi,t \rangle $ with 
$ \mid { \cal {Z}} \rangle $. Therefor
 using
(14), (15) and (16) in  (13), we obtain 
the semiclassical Einstein equation as:
\begin{eqnarray}
{\dot{{\cal{S}}_1}(t)\over {\cal{S}}_1(t)} 
{\dot{{\cal{S}}_2}(t)\over {\cal{S}}_2(t)} + 
{\dot{{\cal{S}}_2}(t)\over {\cal{S}}_2(t)} 
{\dot{{\cal{S}}_3}(t)\over {\cal{S}}_3(t)} + 
{\dot{{\cal{S}}_1}(t)\over {\cal{S}}_1(t)} 
{\dot{{\cal{S}}_3}(t)\over {\cal{S}}_3(t)}&= & 8 \pi \left [ 
\left( \sinh^2 r +{1\over 2} \right ) \left( 
\dot{\phi}^*\dot{\phi}+m^2\phi^* \phi \right )  +   \right. \\ 
\nonumber & & \left.     Re  \left\{\cosh r \sinh r e^{- i 
\varphi} \left ( \dot{\phi}^2 + m^2 \phi^2 \right ) \right\} 
\right ]
\end{eqnarray}

In the above equations, $\phi$ and $\phi^*$ satisfy  the 
boundary condition  
\begin{eqnarray}
 {\cal{S}}_1(t) {\cal{S}}_2(t) {\cal{S}}_3(t)
\left ( \dot{\phi}^*(t) \phi(t)- \phi^*(t) \dot{\phi}(t) \right 
)=i.
\end{eqnarray}

To solve the semiclassical equations (17),  transform the 
solution in the following form
\begin{eqnarray}
\phi(t)=  {\left [{\cal{S}}_1(t) {\cal{S}}_2 (t){\cal{S}}_3(t) 
\right ]}^{- 1/2}\chi(t) 
\end{eqnarray}
Therefor (4) becomes
\begin{eqnarray}
\ddot{\chi}(t)+ \left [ m^2 + { 1 \over 4} \sum_{i=1}^3 \left ( 
{\dot{{\cal{S} }_i}\over {\cal{S}}_i}\right )^2 - {1 \over 2} 
\sum_{i \neq j  =1}^3\left ( { {\dot{{\cal{S}}_i}} (t) \over 
{\cal{S}}_i(t)}{ {\dot{{\cal{S}}_j}} (t) \over {\cal{S}}_j(t)}
 \right ) - { 1 \over 2} \sum_{i=1}^3
{ {\ddot{{\cal{S}}_i}}(t) \over {\cal{S}}_i(t)} \right ] \chi(t)=0.
\end{eqnarray}

The inflaton has a  solution of the form
\begin{eqnarray}
\chi(t)={ 1 \over \sqrt{2 \gamma(t)}}\exp \left ( - i \int 
\gamma(t) dt \right ),
\end{eqnarray}
where,
\begin{eqnarray}
\gamma^2(t)= m^2 + {1 \over 4} \sum_{i=1}^3 \left 
({\dot{{\cal{S}}_i}(t)\over {\cal{S}}_i(t)} \right )^2 -{1 \over 
2} \sum_{i\neq j=1}^3 \left ({\dot{{\cal{S}}_i}(t)\over 
{\cal{S}}_i(t)} {\dot{{\cal{S}}_j}(t)\over {\cal{S}}_j(t)} \right 
) - {1 \over 2} \sum_{i=1}^3 \left ({\ddot{{\cal{S}}_i}(t)\over 
{\cal{S}}_i(t)} \right ) + {3\over4}\left ({\dot{\gamma}(t)\over 
\gamma(t)}\right )^2- {3\over2} {\ddot{\gamma}(t) \over 
\gamma(t)},
\end{eqnarray}
with the following condition:
\begin{eqnarray}
m^2 > { 1 \over 4} \sum_{i=1}^3 \left ( {\dot{{\cal{S}}_i}\over 
{\cal{S}}_i}\right )^2 - {1 \over 2} \sum_{i\neq j=1}^3\left ( { 
{\dot{{\cal{S}}_i}} (t) \over {\cal{S}}_i(t)}{ 
{\dot{{\cal{S}}_j}} (t) \over {\cal{S}}_j(t)}
 \right ) - { 1 \over 2} \sum_{i=1}^3
{ {\ddot{{\cal{S}}_i}}(t) \over {\cal{S}}_i(t)}.
\end{eqnarray}

The  semiclassical equation (17) can be  rewritten  as follows:
\begin{eqnarray}
{\cal{S}}_1(t){\cal{S}}_2(t){\cal{S}}_3(t)&= &    {8 \pi \over 2 
\gamma  \left({\dot{{\cal{S}}_1}(t)\over {\cal{S}}_1(t)} 
{\dot{{\cal{S}}_2}(t)\over {\cal{S}}_2(t)} +
 {\dot{{\cal{S}}_2}(t)\over {\cal{S}}_2(t)}
{\dot{{\cal{S}}_3}(t)\over {\cal{S}}_3(t)} +
 {\dot{{\cal{S}}_1}(t)\over {\cal{S}}_1(t)}
{\dot{{\cal{S}}_3}(t)\over {\cal{S}}_3(t)}\right)} 
 \left [   \left ( \sinh ^2 r + {1 \over 2} \right ) 
\left [  { 1 \over 4 } \sum_{i,j=1}^3 
\left({\dot{{\cal{S}}_i}(t)\over {\cal{S}}_i(t)} 
{\dot{{\cal{S}}_j}(t)\over {\cal{S}}_j(t)} \right) 
\nonumber\right. \right.
\\ \nonumber 
&&\left. \left. + {3\over 4} \sum_{i=1}^3\left( 
{\dot{{\cal{S}}_i}(t)\over {\cal{S}}_i(t)}\right){\dot{\gamma} 
\over \gamma} +  
 {1 \over 4} 
\left ( {\dot{\gamma} \over \gamma}\right )^2 + m^2
 +\gamma^2 \right ] +  
   \left ( \cosh r \sinh r 
  \cos ( \varphi + 2 \gamma t) \right) 
  \right. \\ 
&&\left. \times \left [  { 1 \over 4 } \sum_{i,j=1}^3 
\left({\dot{{\cal{S}}_i}(t)\over {\cal{S}}_i(t)} 
{\dot{{\cal{S}}_j}(t)\over {\cal{S}}_j(t)} \right) +  {3\over 4} 
\sum_{i=1}^3\left( {\dot{{\cal{S}}_i}(t)\over 
{\cal{S}}_i(t)}\right){\dot{\gamma} \over \gamma} + {1 \over 4} 
\left ( {\dot{\gamma} \over \gamma}\right )^2 + m^2
 - \gamma^2 \right ]
 \right ].
\end{eqnarray}

The above equation can be solved peturbatively. Starting from the 
approximation ansatz ${\cal{S}}_{10}(t)={\cal{S}}_{10}t^{n_1}$, 
${\cal{S}}_{20}(t)={\cal{S}}_{20}t^{n_2}$, 
${\cal{S}}_{30}(t)={\cal{S}}_{30}t^{n_3}$, and $\gamma_{0}(t)= m 
$ , we obtain the next order perturbative solution for 
${\cal{S}}_{1}$,
\begin{eqnarray}
{\cal{S}}_{11}(t)&=&
 {8 \pi m\over   {\cal{S}}_{20} {\cal{S}}_{30} \left( 
 n_{1}n_{2}+n_{2}n_{3}+n_{1}n_{3}\right)}
 \left [
 \left( \sinh ^2 r +{1\over 2} \right ) 
 \left(  {1+ {\sum_{i=j=1}^{3}n_{i}n_{j}\over 8 m^2 
t^2}} \right )\right.\nonumber \\
&&\left.   + \cosh r \sinh r    \cos ( \varphi + 2 m t ) \left(  
{{\sum_{i=j=1}^{3}n_{i}n_{j} \over 8 m^2 t^2}-1} \right ) \right 
]  t^{2-n_{2}-n_{3}}.
\end{eqnarray}

Similarly, the next order perturbation solution for 
${\cal{S}}_{2}$ and ${\cal{S}}_3$ are obtained as
\begin{eqnarray}
{\cal{S}}_{22}(t)&=&
 {8 \pi m\over   {\cal{S}}_{10} {\cal{S}}_{30} \left( 
 n_{1}n_{2}+n_{2}n_{3}+n_{1}n_{3}\right)}
 \left [
 \left( \sinh ^2 r +{1\over 2} \right ) 
 \left(  {1+ {\sum_{i=j=1}^{3}n_{i}n_{j}\over 8 m^2 
t^2}} \right )\right.\nonumber \\
&&\left.   + \cosh r \sinh r    \cos ( \varphi + 2 m t ) \left(  
{{\sum_{i=j=1}^{3}n_{i}n_{j} \over 8 m^2 t^2}-1} \right ) \right 
]  t^{2-n_{1}-n_{3}},
\end{eqnarray}
and

\begin{eqnarray}
{\cal{S}}_{33}(t)&=&
 {8 \pi m\over   {\cal{S}}_{10} {\cal{S}}_{20} \left( 
 n_{1}n_{2}+n_{2}n_{3}+n_{1}n_{3}\right)}
 \left [
 \left( \sinh ^2 r +{1\over 2} \right ) 
 \left(  {1+ {\sum_{i=j=1}^{3}n_{i}n_{j}\over 8 m^2 
t^2}} \right )\right.\nonumber \\
&&\left.   + \cosh r \sinh r    \cos ( \varphi + 2 m t ) \left(  
{{\sum_{i=j=1}^{3}n_{i}n_{j} \over 8 m^2 t^2}-1} \right ) \right 
]  t^{2-n_{1}-n_{2}}.
\end{eqnarray}

Where ${\cal{S}}_{11} $ means the next order perturbation solution for the scale 
factor in the x direction and the same hold for ${\cal{S}}_{21}$ 
and ${\cal{S}}_{31} $; respectively in the $ y $ and $ z $ 
directions. 

From (25),(26) and (27) it follow that
\begin{equation}
   ~~~~ {\cal{S}}_{11}\sim t^{2-n_{2}-n_{3}}, ~~~~ 
{\cal{S}}_{22}\sim t^{2-n_{1}-n_{3}}, ~~~~ {\cal{S}}_{33}\sim 
t^{2-n_{1}-n_{2}}.
\end{equation}

Next goal is to examine the particle creation the anisotropically evolving universe described through the Bianchi type I metric. For this, consider the Fock space which has a one parameter dependence on the cosmological time $t$. Then the number of particles at a later time $t$ created from the vacuum at the initial time $t_{0}$ is given by
\begin{equation}\label{1}
    N_{0}(t,t_{0})=\langle 0,\phi,t_{0}| \hat{N}(t)|0,\phi,t_{0}\rangle,
\end{equation}

where $\hat{N}(t)=a^{\dag}a $. Thus using (11), the vacuum expectation value of the right hand side of (29) becomes
\begin{equation}
    \langle \hat{N}(t)\rangle= (S_{1}S_{2}S_{3})^{2} \dot{\phi}\dot{\phi}^{\ast} 
    \langle {\hat{\phi}^{2}}\rangle + \phi \phi^{\ast} \langle\hat{\pi}^{2}\rangle
    - S_{1}S_{2}S_{3}  \phi\dot{\phi}^{\ast} \langle\hat{\pi} \hat{\phi}\rangle -
    S_{1}S_{2}S_{3}\dot{\phi}\phi^{\ast}\langle\hat{\phi}\hat{\pi}\rangle .
\end{equation}
Therefore 
\begin{eqnarray}
 N_{0}(t,t_{0})&=&(S_{1}S_{2}S_{3})^{2}|\phi(t)\dot{\phi}(t_{0})-\dot{\phi}(t) \phi(t_{0})|^{2} .
\end{eqnarray}
The number of particle created in the vacuum states can be obtained by using the peturbative solution in the limit $ m t_{0}$ , $mt > $ 1, in (30), therefore, 
\begin{eqnarray}
  N_{0}(t,t_{0}) &=&  \frac{1}{4\gamma(t)\gamma(t_{0})}
  \frac{S_{1}S_{2}S_{3}}{S_{10}S_{20}S_{30}} 
   \left[ \frac{1}{4} \sum_{i=j=1}^{3} 
  \left( \frac{\dot{S_{i}}\dot{S_{j}}}{S_{i}S_{j}}
  +\frac{\dot{S_{i0}}\dot{S_{j0}}}{S_{i0}S_{j0}} \right)
  - \frac{1}{2}\sum_{i=j=1}^{3}\frac{\dot{S_{i}}\dot{S_{j0}}}{\dot{S_{i}}\dot{S_{j0}}}  
  \right.\nonumber\\
 &&\left. +\frac{1}{2}\sum_{i=1}^{3}\frac{\dot{S_{i}}}{S_{i}} \left(
  \frac{\dot{\gamma}(t)}{\gamma(t)}+\frac{\dot{\gamma}(t_{0})}{\gamma(t_{0})}
  \right) 
    +\frac{1}{2}\sum_{i=1}^{3}\frac{\dot{S_{i0}}}{S_{i0}} \left(
  \frac{\dot{\gamma}(t)}{\gamma(t)}-\frac{\dot{\gamma}(t_{0})}{\gamma(t_{0})}
  \right) \right.\nonumber \\ 
 && \left.+
  \left[\frac{1}{2}\left(\frac{\dot{\gamma}(t)}{\gamma(t)}-
  \frac{\dot{\gamma}(t_{0})}{\gamma(t_{0})}\right)\right]^{2}
  + \gamma(t)^{2}-\gamma(t_{0})^{2}
  \right]  \nonumber \\
  &\simeq &\left(\frac{n_{1}+n_{2}+n_{3}}{4 m}\right) ^{2}
  \left(\frac{t-t_{0}}{tt_{0}} \right)^{2}
  \left(\frac{t}{t_{0}}\right)^{n_{1}+n_{2}+n_{3}} .
\end{eqnarray}
By similar procedure, one can compute the particle creation 
for the quantized inflaton in squeezed vacuum states also.
For this, consider the quantized inflaton in the squeezed vacuum state formalism.
 Then the expectation values of 
$ \pi^{2}, \phi^{2}, \pi\phi$ and  $\phi\pi$
 can be computed in squeezed vacuum state  by using (11),(14) and (16), and are respectively obtained as

\begin{eqnarray}
 \langle \hat{\pi}^{2} \rangle _{sqv}&=& (S_{1}S_{2}S_{3})^{2}
 \left[\left(2 \sinh^{2} r + 1\right)\dot{\phi}^{\ast}(t_{0})\dot{\phi} 
 (t_{0})
 + \sinh r \cosh r \left( e^{-i\varphi}\dot{\phi}^{\ast 2 }(t_{0})
 + e^{i\varphi}\dot{\phi}^{ 2 }(t_{0})\right)\right] \nonumber\\
  \langle \hat{\phi}^{2} \rangle _{sqv} &=& 
  \left(2 \sinh^{2} r + 1\right){\phi}^{\ast}(t_{0}){\phi} (t_{0})
 + \sinh r \cosh r \left( e^{-i\varphi}{\phi}^{\ast 2 }(t_{0})
 + e^{i\varphi}{\phi}^{ 2 }(t_{0})\right)
\nonumber\\
 \langle \hat{\pi} \hat{\phi}\rangle _{sqv}&=& S_{1}S_{2}S_{3}\left[
\sinh^{2} r \dot{\phi}^{\ast}(t_{0})\phi (t_{0}) +
 \cosh^{2} r \dot{\phi}(t_{0})\phi^{\ast} (t_{0})+\right. 
 \nonumber \\
&&\left. \sinh r \cosh r \left( e^{-i\varphi}\dot{\phi}^{\ast  
}(t_{0})\phi (t_{0})
 + e^{i\varphi}\dot{\phi}(t_{0})\phi (t_{0})\right)\right]
\\
 \langle \hat{\phi}\hat{\pi} \rangle _{sqv}&=& S_{1}S_{2}S_{3}\left[
\sinh^{2} r {\phi}^{\ast}(t_{0})\dot{\phi }(t_{0}) +
 \cosh^{2} r {\phi}(t_{0})\dot{\phi}^{\ast} (t_{0})+ 
 \right. 
 \nonumber \\
&&\left.
 \sinh r \cosh r \left( 
e^{-i\varphi}{\phi}(t_{0})\dot{\phi } ^{\ast}(t_{0})
 + e^{i\varphi}{\phi}(t_{0})\dot{\phi} (t_{0})\right)\right].\nonumber
\end{eqnarray}
Applying the above results (33) in (30), the number of particle created in squeezed vacuum in the limit $ m t_{0}$, $mt > $ 1, is obtained as

\begin{eqnarray}
  N_{sqv}(t,t_{0}) &=& \frac{1}{4 \gamma (t)\gamma(t_{0})}
   \frac{S_{1}S_{2}S_{3}}{S_{10}S_{20}S_{30}}
    \left[ \left(2 \sinh ^2 r +1 \right) \left(
    \frac{1}{4}\left(\sum_{i=1}^{3}\frac{\dot{S}_{i}(t_{0})}{S_{i}(t_{0})}+
    \frac{\dot{\gamma}(t_{0})}{\gamma(t_{0})}\right)^{2} + \gamma^{2}(t_{0})+\gamma^{2}(t)
    \right. \right. \nonumber\\
  && \left.  \left.+\frac{1}{4}\left(\sum_{i=1}^{3}\frac{\dot{S}_{i}(t)}{S_{i}(t)}+
    \frac{\dot{\gamma}(t)}{\gamma(t)} \right)^{2}  \right)+ \sinh r \cosh r \left(
    e^{-i(\varphi- 2 \int\gamma(t_{0})dt_{0})}-
    e^{i(\varphi- 2 \int\gamma(t_{0})dt_{0})}\right)\right.  \nonumber \\
   && \left. \times \left( \frac{1}{4}\left(\sum_{i=1}^{3}
   \frac{\dot{S}_{i}(t_{0})}{S_{i}(t_{0})}+
    \frac{\dot{\gamma}(t_{0})}{\gamma(t_{0})}\right)^{2}+
    \frac{1}{4}\left(\sum_{i=1}^{3}\frac{\dot{S}_{i}(t)}{S_{i}(t)}+
    \frac{\dot{\gamma}(t)}{\gamma(t)} \right)^{2} +\gamma^{2}(t)  - 
    \gamma^{2}(t_{0})
    \right)\right. \nonumber \\
   &&\left.- \sinh^2 r   \left( \frac{1}{2}\left(\sum_{i=1}^{3}\frac{\dot{S}_{i}(t)}{S_{i}(t)}+
    \frac{\dot{\gamma}(t)}{\gamma(t)} \right) 
    \left(\sum_{i=1}^{3}\frac{\dot{S}_{i0}(t)}{S_{i0}(t)}+
    \frac{\dot{\gamma}(t_{0})}{\gamma(t_{0})} \right) -2\gamma(t)\gamma(t_{0})  
    \right) 
    \right. \\ \nonumber
    &&\left.-\cosh^2 r
     \left( \frac{1}{2}
   \left(\sum_{i=1}^{3}\frac{\dot{S}_{i}(t)}{S_{i}(t)}+
    \frac{\dot{\gamma}(t)}{\gamma(t)} \right)  
    \left(\sum_{i=1}^{3}\frac{\dot{S}_{i0}(t)}{S_{i0}(t)}+
    \frac{\dot{\gamma}(t_{0})}{\gamma(t_{0})} \right) +2 \gamma(t)\gamma(t_{0})  
    \right) 
    \right.  \nonumber\\
    &&\left.- \sinh r \cosh r  \left(e^{-i\varphi}+
     e^{i(\varphi-2 \int\gamma(t_{0})dt_{0})}\right)
    \left( \frac{1}{2}
  \sum_{i=1}^{3}\frac{\dot{S}_{i}(t)}{S_{i}(t)}+
    \frac{\dot{\gamma}(t)}{\gamma(t)} \right) 
    \left(\sum_{i=1}^{3}\frac{\dot{S}_{i0}(t)}{S_{i0}(t)}+
    \frac{\dot{\gamma}(t_{0})}{\gamma(t_{0})} 
     \right)  
    \right].  \nonumber 
\end{eqnarray}    
Again using the peturbative solution , we get 
   \begin{eqnarray}
  N_{sqv}(t,t_{0}) 
  &\simeq & 
  \frac{\left(n_{1}+n_{2}+n_{3}\right)^{2}}{4m^{2}}
  \left( \frac{t}{t_{0}}\right)^{n_{1}+n_{2}+n_{3}}
   \left[ \left(2 \sinh^{2} r + 1 \right)\left(\frac{t-t_{0}}{2t_{0}t}\right)^{2}
  \right. \nonumber \\ 
&& \left. 
  +4 m^{2}\sinh ^{2}r- \frac{\sinh r \cosh r \cos \varphi}{2 t t_{0}} \right].
  \end{eqnarray} 
Therefor the total number of particle created for the whole range of squeezing parameter and squeezing angle can be written as  
\begin{eqnarray}
     N_{sqv-tot}= N_{0}\left(t,t_{0}\right)+  \int_{0}^{r_{max}} N_{sqv-r}  dr + 
     \int_{0}^{\varphi_{max}} N_{sqv-\varphi} d \varphi ,
  \end{eqnarray}   
  
  where
  \begin{eqnarray}
N_{sqv-r}   &=& 
    \frac{\left(n_{1}+n_{2}+n_{3}\right)^{2}}{4m^{2}}
  \left[ 4  \sinh r \cosh r  \left(\left(\frac{t-t_{0}}{2t_{0}t}\right)^{2}+2 m^{2}\right)
  \right. \nonumber \\
 && \left.
  - \frac{\left(\sinh^{2} r+ \cosh^{2}r \right)\cos \varphi}{2 t t_{0}} \right]
  \left( \frac{t}{t_{0}}\right)^{n_{1}+n_{2}+n_{3}}, \\ \nonumber
  \end{eqnarray}  
 and 
  \begin{eqnarray}
 N_{sqv-\varphi}  &=& 
   - \frac{\left(n_{1}+n_{2}+n_{3}\right)^{2}}{4m^{2}}
  \left[ 
   \frac{\sinh r \cosh r \cos \varphi}{2 t t_{0}} \right]
  \left( \frac{t}{t_{0}}\right)^{n_{1}+n_{2}+n_{3}}. \\ \nonumber
  \end{eqnarray}

{\bf 4. Conclusions}

In this paper we have examined the behavior a homogeneous and 
massive scalar (inflaton) field minimally coupled to the gravity in  Bianchi 
type -I model of the universe, in the frame work of semicalssical 
theory of gravity. The inflaton is represented in squeezed vacuum state formalism of quantum optics and hence the approximate leading solution to the 
semiclassical Einstein equation  is found. The next order solution 
for each scale factor in their respective direction show that 
each scale factor in each direction dependent on 
power law of expansion. Further more the solutions  show that evolution of scale 
factors are mutually correlated.
 When $n_1 = n_2=
n_3 = n $, then the corresponding solution 
reduces to  isotropic model  and is consistent with the result 
obtained as in the ref. 23. Form anisotropic to isotropic 
transition a damping mechanism is required.  One of the efficient 
damping mechanisms could  be due to the particle creation in  
anisotropic models.  We have also examined the nonclassical particle creation in Bianchi type -I cosmological model by representing the inflaton in squeezed vacuum state formalism. 
The present study can  account for the nonclassical particle creation and power law of expansion of 
the scale factors in Bianchi type -I universe, for a homogeneous 
and massive scalar field minimally coupled to the gravity,in the 
frame work of semiclassical theory of gravity.

{\bf Acknowledgements}

P.K.S. wishes to thank the Director and IUCAA, Pune, for warm 
hospitality and library facilities made available to him and 
acknowledges Associateship of IUCAA.

\begin{center}
{\bf References}\end{center}

1.C.W.Misner,{\it Phys.Rev.}{\bf 186},1328(1969)\\
2.C.W.Misner, K.S.Thorne and J.A.Wheeler,\emph{Gravitation }(Freeman,San Fransico)(1973)\\
3.O.Heckman and E.Schucking,
{\it Gravitation : An Introduction to Current Research }(ed.L.Witten)(Wiley,New York,1962) \\
4.K.S.Throne, {\it Ap.J.} {\bf 148},51 (1967)  \\
5.V.A.Belinski, E.M.Lifshitz, and I.M.Khalatnikov,
{\it Sov.Phys.Usp.} {\bf 13},745 (1971)\\
6.Ya.B.Zel'dovich and A.A.Starobinsky,
{\it Zh.Eksp.Teor.Fiz.} {\bf 61},2161(1971)\\
7.B.L.Hu and L.Parker,
{\it Phys.Rev.} {\bf D17},933,(1978)\\
8.W.H.Huang, {\it Phys.Rev.} {\bf D42}, 1287 (1990).\\
9.F.Futamas, {\it Phys.Rev.} {\bf D29}, 2783 (1984).\\
10.A.L.Berkin, {\it Phys.Rev.} {\bf D46}, 1551 (1992).\\
11.V.A.Belinski and I.M.Khalatnikov,
   {\it Sov.Phys.JETP.} {\bf 63},1121 (1972)\\
12.V.N.Folomeev and V.Ts.Gurovich, {\it Gen.Rel.Grav.} {\bf 
32},1255(2000)\\ 
13.B.L.Shumaker,\emph{Phys.Rep}.{\bf 135}317(1986)\\
14.L.P.Grishchuk and Y. V. Sidorov, {\it Phys.Rev.} {\bf D42}, 3413 (1990).\\
15.R.Brandenberger, V. Mukhanov and T.Prokopec, {\it Phys.Rev.Lett.} {\bf 69}, 3606 (1992).\\
16.C.Kuo and L.H.Ford, {\it Phys.Rev.} {\bf D46}, 4510 (1993).\\
17.A.Albrecht .; et.al. {\it Phys.Rev.} {\bf D50}, 4807 (1994).\\
18.M.Gasperini and M.Giovananni, {\it Phys.Lett.} {\bf B301}, 334 (1993).\\ 
19.B.L.Hu, G.Kang and A.L.Matacz, {\it Int.J.Mod.Phys.} {\bf A9}, 991 (1994).\\
20.P.K.Suresh, V.C.Kuriakose, and K.Babu Joseph, {\it Int.J.Mod.Phys.} {\bf D6}, 771 (1995).\\
21.P.K.Suresh and V.C.Kuriakose, {\it Mod.Phys.Lett.} {\bf A13}, 165 (1998).\\
22.P.K.Suresh, {\it Mod.Phys.Lett.} {\bf A16}, 2431 (2001).\\
23.S.P.Kim and D.N.Page, {\it J. Korian.Phy.Soc} {\bf 35}, S660 (1999).\\
\end{document}